\documentclass[11pt]{article}

\usepackage{amssymb}
\usepackage{amsmath}
\usepackage{bm}

\usepackage{graphicx}
\usepackage{epstopdf}
\begin{document}

\title{The Early History of the Aharonov-Bohm Effect.}
\author{B. J. Hiley\footnote{E-mail address b.hiley@bbk.ac.uk.} }
\date{TPRU, Birkbeck, University of London, Malet Street, \\London WC1E 7HX. \vspace{0.4cm} }
\maketitle

\begin{abstract}
This paper traces the early history of the Aharonov-Bohm effect.  It appears  to have been `discovered' at least three times to my knowledge before the defining paper of Aharonov and Bohm appeared in 1959.   The first hint of the effect appears in Germany in 1939, immediately disappearing from sight in those troubled times.  It reappeared in a paper in 1949, ten years before the defining paper appeared.  Here I report the background to the early evolution of this effect, presenting first hand unpublished accounts reported to me by colleagues at Birkbeck College in the University of London. 
\end{abstract}

\section{Introduction.}

My interest in the history of the Aharonov-Bohm effect [AB] started when I joined Birkbeck College  in 1961 where two of key players in the discovery held professorships.  Werner Ehrenberg held the Chair of Experimental Physics and was head of the Department, while Bohm had just been appointed to a Chair in Theoretical Physics.  My first venture into this effect resulted in some personal embarrassment.  I had decided to write a paper on the effect mainly to clarify my own understanding of the phenomenon.  When Ehrenberg saw a copy of the paper he confronted me with a comment spoken with a strong German accent, ``Ach Hiley, zis AB effect that you are discussing, is it the one that Siday and I discovered?"  Poor dear, gentle Werner, sidelined in history by a young new member of his own staff!

It was my original intention to describe the way in which Ehrenberg and Siday \cite{wers} discovered the effect in the 1940s.  I had at hand in the Department a number of people who were working with Ehrenberg and Siday when they made their discovery, so I am able to obtain first hand accounts of how the story unfolded.  My only regret was that Ray Siday had passed on, but the first hand recollections of the man were still very much alive in the department.  The events leading up to the discovery took place in the late 1940s and it turns out to be a fascinating story. 

 I will only briefly comment on the re-discovery  of the effect by Yakir Aharonov and David Bohm \cite{yadb59} in the late 1950s.  The brevity of my comments is no reflection on their work, which was very significant since it marked the beginning of gauge field theories \cite{mpat89}.  Although I had many hours of discussion with David Bohm, the $AB$ effect was not high on the agenda, as we were interested in a wider range of physical and philosophical questions.  I will leave a discussion of the AB's papers and the subsequent to the development of general gauge theory to others.

 \section{Did Walter Franz discover the $AB$ effect in 1939?}
 
 As I was beginning to collect the background material, my attention was drawn to the abstract of a talk given by Werner Franz to a physical society meeting in Danzig in 1939 \cite{wf39}.  It left me with the impression he might have been aware of the effect  even at this early date.  Could the talk he presented have contained a first mention of this effect and {\em ipso facto} could he have been the first to discover this effect?
 
 How does one investigate something that took place in the then free city of Danzig in May 1939, when it was just about to be involved in violent political events?.  By August, the city had undergone a {\em coup d'etat} and later in September a German battleship used the harbour to open fire on the Polish city of Westerplatte. Not the atmosphere to think deeply about physics!
 
 The starting point for this investigation, then, can only be the Franz abstract.   Here we reproduce it in its original form.
 
\begin{quote}
Item 5. Hr. W. Franz (K\"{o}nigsberg, Pr): {\em Elektroneninterferenzen im Magnetfeld.} 

Nach de Broglie ist einem Elektron vom Impuls $p$ eine Wellenl\"{a}nge $\lambda  = h/|p|$ zugeordnet, worin $h$ die Plancksche Konstante. Im Magnetfeld tritt zum korpuskularen Impulse $mv$ der Zusatz $eA/c$, wo $A$ das Vektor potential des Magnetfeldes.

Da $A$ nur bis auf einen Gradienten bestimmt ist, hat im Magnetfeld auch die Wellenl\"{a}nge keinen eindeutigen physikalischen Sinn. Doch f\"{a}llt diese Vieldeutigkeit bei der physikalischen Anwendung, der Interferenz. heraus. - Der Wellenzahlvektor eines Eleklronenfeldes 
$k = 2\pi /\lambda $., nach de Broglie also $p/h$ ist der Gradient der Phase, muss  also rot-frei sein. Im Magnetfeld ist nun rot $(mv) \ne 0$; der Zusatz $eA/c$ stellt die Bedingung rot $k = 0$ wieder her. - Bei der Beugung am Doppelspalt ergibt sich als Bedingung f\"{u}r ein Interferenzmaximum $mr\Delta s + e\Phi/c =nh$ 
worin $\Delta s$ die geometrische Gangdifferenz der beiden m\"{o}glichen
korpuskularen Bahnen und $\Phi$ der magnetische Fluss durch die von ihnen eingeschlossene Fl\"{a}che ist $n$ eine ganze Zahl. Hiernach ergibt sich in \"{U}bereinstimmung  mit der Erfahrung, dass  die Interferenz nur durch die Richtung bestimmt wird, in welcher die  Elektronen die Spalte erreichen und verlassen. Die durch die Bahnkr\"{u}mmung im Magnetfeld hervorgerufenen Gangdifferenzen werden durch den Zusatz $e\Phi /c$ aufgehoben\footnote{A colleague, John Jennings , who had worked in RAF Intelligence with R. V. Jones during the World War II, kindly translated this as:-
 \begin{quote}
 According to de Broglie, a wave length $\lambda = h/|p|$ is associated with an electron of momentum $p$, where $h$ is  Planck's constant.  In a magnetic field, $ eA/c$ is added to the particle's momentum $mv$, where $A$ is the vector potential of the magnetic field.  Since $A$ is undefined to within an added gradient, the wave length has not a unique physical meaning in a magnetic field.  However, this ambiguity is removed by the application of interference.  The wave number vector  $k = (2\pi /\lambda)$ of an electron field (which according to de Broglie is $p/h$) is then the gradient of the phase and must therefore be irrotational.  In a magnetic field curl($mv) \ne 0$; the additional term  $ eA/c$ again produces the condition curl $k = 0$.  On diffraction at a double slit, the condition for an interference maximum is $mv\Delta s + e\Phi/c = nh$, where $\Delta s$ is the geometric path difference of the two possible particle paths and $\Phi$ is the magnetic flux through the surface enclosed by them, and $n$ is an integer.  In agreement with experiment, this shows that the interference is determined only by the directions in which the electrons reach and leave the slits.  The path difference caused by the curvature of the paths in the magnetic field falls out, on account of the additional term $ e \Phi/c$ \cite{jj89}
\end{quote}}
\end{quote}

Although Franz clearly states that the interference depends on the magnetic flux enclosed by the electron paths, he adds that ``the path difference is caused by the curvature of the paths in  the magnetic field".   This last statement clouds the issue because it is unclear whether he realised that the electron paths need not experience any magnetic field at all and still produce the same interference shift provided the path encloses the field.  This is the key to the whole effect.

I originally thought that this was an abstract of a paper, but I failed to find any such paper.  It then transpired that it was an abstract of a seminar that Franz gave to the physics meeting ``Gauvereins Ostland in Danzig" held from 18 to 29 May   1939.

 Soon after this abstract came to my attention \cite{op85}, I was approached by Gottfried M\"{o}llenstedt, an experimentalist who had pioneered some brilliant electron interference experiments \cite{gm56}, and asked to supply some biographical details of both Ehrenberg and Sidday.  He was preparing a paper on the history of the electron biprism \cite{gm98}, an instrument that he had perfected and had used in his electron of interference experiments.
 
In this article, M\"{o}llenstedt \cite{gm98} informs us that he attended the Danzig meeting as a young physics student and it was at that meeting that  he heard the term ``electron interferometry" for the first time. To recall the meeting he used a paper by Franz \cite{wf65}, which was written in 1965, and translates a key part of this paper which is reproduced here.
\begin{quote} 

In presence of an electromagnetic field, the momentum of a particle with charge $e$ is known to be
\begin{eqnarray*}
\overrightarrow p=m\overrightarrow v+e\overrightarrow A 
\end{eqnarray*}								
where $\overrightarrow A$ is the vector potential from which the magnetic field strength is determined as $\overrightarrow B =\nabla \times \overrightarrow A$. The phase difference $\Delta \phi $ between two rays (a) and (b) connecting two points 1 and 2 is determined by 
\begin{eqnarray*}
\Delta \phi=\int_{1(a)}^{2(a)}\overrightarrow p d \overrightarrow r +\int_{1(b)}^{2(b)}\overrightarrow p d \overrightarrow r = \oint \overrightarrow p d \overrightarrow r
\end{eqnarray*} 									
Introducing the expression for $\overrightarrow p $ is from above, the term $m\overrightarrow v$ yields the same path difference as in absence of a magnetic field whereas, according to Stokes' theorem, the loop integral over $\overrightarrow A$ may be transformed to a surface integral over $\nabla \times \overrightarrow A$, i.e., the magnetic flux $\Phi_{m}$, yielding 
\begin{eqnarray*}
\Delta \phi = \oint m\overrightarrow v d\overrightarrow r + e\Phi_{m}
\end{eqnarray*}									
This simple relation which should be the first thing taught in a lecture on wave mechanics for beginners after introducing the magnetic field (strangely enough I could not find it in any lecture notes except my own) shows that the phase difference between electron rays depends on the magnetic flux included between the rays, even if the rays do not run in a magnetic field. 
\end{quote}
 In 1939, after the lecture by Walter Franz on, Walther Kossel discussed the possibility of an 
experimental proof, but he came to the conclusion that, at that time, an experimental proof was not feasible.
 M\"{o}llenstedt writes,
\begin{quote}
I remember W. Kossel saying  ``I hear the message but I lack an electron 
interferometer."  
\end{quote}

My reading of these last two sentences strongly suggests that there was a discussion about the AB effect at that Danzig meeting, but it was brought to a end simply because no one in the group had the means to explore the effect experimentally.  But did that mean that Franz  had really discovered the effect?

In contrast to this, I have a letter from M\"{o}llenstedt dated Feb. 12 1993, which is six years before M\"{o}llenstedt's paper appeared \cite{gm98}. In it M\"{o}llenstedt writes

\begin{quote}
I am enclosing a copy of a recent paper on the ``Amerigo effect" printed in ``Physikalische Bl\"{a}tter". I think that possibly the AB effect is also some kind of an Amerigo effect for which the credit should rather have been given to Ehrenberg and Siday. 
\end{quote}

The ``Amerigo effect" was a phrase coined by M\"{o}llenstedt and Walther Kossel \cite{km93} to refer to the  mis-attribution of the discoverer of some physical effect.  The word takes its meaning from Amerigo Vespucci who sailed West after Columbus made his discovery of America.  The Medici Bank of Florence, for whom Amerigo worked as a local branch manager in Seville, spotting an advertising opportunity, decided to give circulation in Europe to an account of Amerigo's journey.  A geographer Waldseem\"{u}ller subsequently attributed the discovery of the new continent to Amerigo!

Unfortunately M\"{o}llenstedt died in 1997 and by the time the ``History of the Electron Biprism" article reached me, it was too late  to obtain a clarification of this point, so it is still an open question as to whether Franz had spotted the effect.

 \section{Some Background to the Work of Ehrenberg and Siday.}
 
Werner Ehrenberg was born in Berlin and studied Philosophy, Physics and Mathematics at the University of Berlin and took his PhD at Heidelberg. He was an Assistant in the Physical Institute of the Technische Hochschule in Stuttgart but was dismissed for being Jewish in 1933 and sort refuge in the UK.  He worked at Birkbeck under Blackett on a grant from the Academic Assistance Council, before working in industry during WW II.  At the end of the war he returned to Birkbeck where he worked under J. D. Bernal before becoming established in his own right, finally taking the Chair of Experimental Physics at Birkbeck.
 
His interests in physics were wide and varied.  His early interests were in x-rays and electron optics. Indeed this early work centred on the task of developing a method to produce soft focus x-rays that could be used on biological molecules \cite{WEAF}, \cite{WEWS}. With Walter Spear he developed and built the fine focus X-ray generator that Maurice Wilkins, Rosalind Franklin and Ray Gosling used to study the structure of DNA.  It was this experimental work that enabled Crick and Watson to propose the double helix structure that won them, together with Wilkins, the Nobel prize. Ehrenberg's later interests were in electrical conduction in semiconductors and metals \cite{WESC}.
 
Ray Siday was a totally different character.  He took a First in the B.Sc. Special Physics (London) and worked with Patrick Blackett on Nuclear physics.  In 1938 after spending several years on the South Sea Island of Tahiti, he returned to Birkbeck and began working on beta-spectra.  Here he developed a keen interest in electron optics.  At the time of the publication of their paper reporting what has become known as the AB effect, he was working in Edinburgh University.
  
The collaboration between Ehrenberg and Siday started in 1933 when both of them were first at Birkbeck, although from reading  the personal recollections of Werner Ehrenberg, it seems that many of these discussions took place in the local pubs!  These early discussions were eventually interrupted by Ray Siday's adventures in Tahiti and then, of course, by the war, so, in effect, they did not start collaborating again until after the war when they both returned once again to Birkbeck. 

 Their discussions were very wide ranging, but it was the principles of electron optics that  was the focus of their attention and it was these discussions that led them eventually to predict what has become known as the AB effect.  Their paper  reporting this effect \cite{wers} appeared ten years before the classic paper of Aharonov and Bohm \cite{yadb59}. I say at least, because in 1946, Siday took a three year ICI fellowship at Edinburgh to continue his work on beta-spectra with Norman Feather who, in turn, had previously worked with Ernest Rutherford in the Cavendish at Cambridge.
 
Siday's work on beta spectra involved, among other things, the focussing of the beta rays in magnetic fields.  This is what sustained his  interests in electron optics in general.  In the pioneering days of designing electromagnetic lenses, much use was made of the analogy with optical lens systems. Of course in optics, one had long been aware of the tensions between ray optics on the one hand and wave optics on the other.  In the case of electrons there was a similar tension, between the particle properties, rays, and their wave properties.

 In geometric optics the equation of the ray between two points $P_{1}$ and $P_{2}$ is obtained from the variation principle,
\begin{eqnarray*}
\delta\int_{P_{1}}^{P_{2}}\mu ds=0
\end{eqnarray*}
This is essentially Fermat's principle which is analogous to Hamilton's principle in dynamics
\begin{eqnarray*}
\delta\int_{P_{1}}^{P_{2}}p ds=0
\end{eqnarray*}
Here $p$ is the conjugate momentum obtained from the Lagrangian of the electron in an electromagnetic field.  Ehrenberg and Siday showed, in a manner that would be regarded as rather clumsy today, that this momentum would be given by
\begin{eqnarray*}
\delta\int_{P_{1}}^{P_{2}}[m{\bm v.}{\bm dr}+{\bm A.}{\bm dr}+\phi dt]
\end{eqnarray*}
Thus they concluded that one could think of a refractive index for the electron lens as given by
\begin{eqnarray}
\int_{P_1}^{P_2} \mu ds=\int_{P_1}^{P_2} [m{\bm v.}{\bm dr}+{\bm A.}{\bm dr}+\phi dt]   \label{E:ref}
\end{eqnarray}
Here $\int_{P_{1}}^{P_{2}}\mu ds$ is simply the optical path length of the ray, so that if we divide this by the de Broglie wave length of the electrons, we can find the phase difference  between the points $P_{1}$ and $P_{2}$ on the ray. 

The practical problem that Siday was thinking about was the design of a magnetic lens for his beta spectrometer, so the electrostatic potential $\phi=0$ was put to zero.  Thus one considered an optical path defined by $\int [m{\bm v.}{\bm dr}+{\bm A.}{\bm dr}]$.  Now one can clearly see a problem.  This optical path depends upon the vector potential $\bm A$, so clearly the refractive index is not a gauge invariant expression.  This problem was clearly recognised by Ehrenberg and Siday and is discussed carefully in their paper.

In order to motivate the discussion of gauge invariance, Ehrenberg and Siday recalled the mathematical conditions that must be placed on an optical refractive index. This index, $\mu$, is a measured quantity and so must be finite and single valued.  It must also be continuous, except at a finite number of surfaces separating any different media traversed by the ray.  

The same conditions must be satisfied by the equivalent electron refractive index defined by the RHS of equation \eqref{E:ref}.  This means that the refractive index must be fixed everywhere in space once it is fixed in the neighbourhood of one point.  Furthermore, it must be single valued.  It should have no singularities and any discontinuities should be of such a nature that they appear as limiting cases of a continuous refractive index.  

Since ${\bm A.}{\bm dr}$ occurs as an additive term in the refractive index, the same conditions must be applied to it and hence to ${\bm A}$ itself.  Now these conditions are just those for the validity of Stokes theorem and this is the only valid restriction which must be imposed on it.  A consequence of this is that ${\bm A}$ cannot in general be chosen so as to vanish even though the magnetic field vanishes locally. It is this condition that is vital for understanding the AB effect.

As is well known by now, the ambiguity in ${\bm A}$ arises because all we insist on is that $\nabla\times {\bm A}={\bm B}$, the magnetic field.  Thus it is always possible to add to ${\bm A}$ the gradient of a scalar, $\nabla \psi$, without changing ${\bm B}$.  This follows directly from the identity $\nabla\times\nabla =0$.  This arbitrariness cannot produce any observable effects in the geometrical aspects of the optics.  

However wave optics is a different matter.  To bring out the consequences, Ehrenberg and Siday consider the phase difference $d$ between any two paths rays.  Using \eqref{E:ref} we have
\begin{eqnarray*}
d=\int _{0}^{1}[m{\bm v}+{\bm A}]{\bm dr}-\int _{0}^{1}[m{\bm v}+{\bm A}]{\bm dr'}
\end{eqnarray*}
where ${\bm dr}$ is an element of the first path and ${\bm dr'}$ is an element of the second path.  Now since the momentum of the electron is constant and using Stokes theorem, we find
\begin{eqnarray*}
d=\frac{h}{\lambda}\left(l_{1}-l_{0}\right)+\iint \bm B.\bm d{\boldsymbol \sigma}.
\end{eqnarray*}
Thus we see that, again as in the case of Franz, the phase difference between the two paths depends on the flux enclosed by the closed path.

Now Ehrenberg and Siday take the argument one stage further. They  consider the specific case when the magnetic field is confined to a region within the circuit in such a way that the electrons do not pass through the magnetic field.  In other words, if the electron should follow either path it will not experience any field.  They then ask the crucial question: ``Can we gauge transform away the vector potential so that there is no contribution from the magnetic flux lying entirely within the circuit?"  They show that this is not possible without introducing a singularity which would lead to the  breakdown of Stokes theorem.

\begin{figure}[h] 
   \centering
   \includegraphics[width=4in]{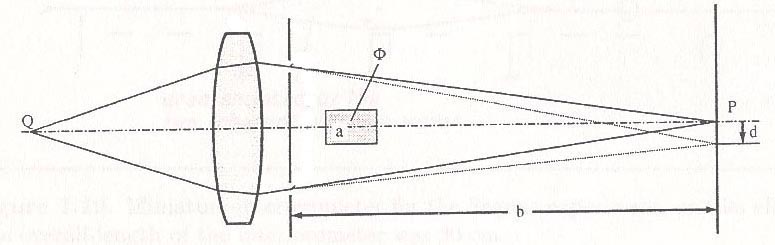} 
   \caption{Magnetic flux $\Phi$ isolated from rays}
   \label{fig:MagFlux}
\end{figure}

Thus they conclude that it is not possible to find a vector potential which satisfies Stokes' theorem and removes the anisotropy of the whole space outside the field.  The fact that this irremovable anisotropy from the field free region as a whole emphasises the fact that the electron-optic refractive index contains the vector potential and not the magnetic field strength.  This led them to the conclusion that wave-optical effects will arise from an isolated magnetic field even though the rays travel in a field free region.  To emphasise this claim they sketch an experimental situation that would demonstrate this effect. This is shown in figure \ref{fig:MagFlux}

Both Ehrenberg and Siday found this result very strange and totally contrary to what they would expect.   At that stage the vector potential ${\bm A}$ was simply regarded as a mathematical symbol with no observable consequences.  Yet here, in quantum mechanics, it has an observable consequence.    Siday was overheard by John Jennings \cite{jj89} one day remarking to Ehrenberg, ``Isn't it odd that the only physical effect produced by a magnetic flux is in a superconductor?"  Ehrenberg replied ``Ach! but what better superconductor than free space!"

Siday went around explaining the effect to anyone who would listen and asking where the argument had gone wrong, if indeed it had.  All were puzzled and felt it was wrong, but could not put their finger on the error.  Siday became incensed with the lack of enthusiasm.  ``There are all these sodding geniuses poncing around but if they ever have to do anything- oh yes, that's a different matter isn't it!"  Eventually he became so frustrated that he decided to present the conundrum to Max Born to see what he would say.

Recall that  at this time, Siday was working at Edinburgh University in Norman Fowler's laboratory, while Born was then Tait professor of Natural Philosophy at the same university,  so he decided to invite Max Born to his laboratory.     David Butt\footnote{David Butt, Emeritus Reader in Experimental Physics, Birkbeck,  was a research student in Edinburgh at the time working on internal conversion of $\beta$-rays.  Butt was one of the Birkbeck group that carried out one of the first experiments to test for quantum non-locality holding over distances of up to 6m. \cite{wlb75}.} shared the same laboratory with Siday and at the time of the meeting, was sitting in the corner of the lab while Siday and Born discussed the effect.  Unfortunately he could not hear the actual conversation, which lasted about 45 minutes, but he did see Born shaking his head from side to side every so often, seemingly with incomprehension.   

At the end of the discussion, the two of them rose, shook hands and Born departed with a face looking like thunder.  As soon as the door closed, Siday came up to Butt and exclaimed ``Well, sod it, I got absolutely nothing out of him.  Who can I ask now?" \cite{db99}.
  Butt has often puzzled about Born's  reaction.  The problem was simple enough to describe.  All the others, mainly experimental colleagues, had understood what Siday was claiming, but could not spot the `error.'  If what Siday had presented to Born was an obvious consequence of standard quantum mechanics, surely Born would have said so, but he didn't.  Nor did he offer any criticism of the explanation of the effect.

In spite of this apparent indifference, Ehrenberg and Siday decided to go ahead and publish the work in the Proceedings of the Physical Society, which was one of the main British journals at the time.  Unfortunately they clearly had not been advised by a publicity agent as the paper was entitled ``The Refractive Index in Electron Optics and the Principles of Dynamics", a title which gives no clue as to the radical nature of what they had discovered.  Even the abstract did not highlight the effect they had found, but the conclusion was quite clear.  One should observe a fringe shift proportional to the magnetic flux included and isolated from the passing electrons.

David Butt also informed me that a few years later in about 1957, he had been talking with Siday, and asked him if either he or Ehrenberg had been contacted by anyone concerning the paper.  Siday's is reported as saying, ``No.  We have not heard a bloody thing--not as much as a whisper.  It has fallen to the bottom like a lump of lead". 

But why should they?  At that stage, the vector potential $\bm A$ was still regarded as merely a mathematical convenience and could be gauge transformed away.  Therefore it should produce no physical effect.  Furthermore  the effect was presented in a context that it appeared to be a problem in designing electron lenses, not a general new effect.  The choice of title of their paper only confirmed this.   However it was not only the title of the paper, the presentation suffered from two further disadvantages. 

 Firstly the effect was discussed in specialist terms of `equivalent refractive indices', using the optical path analogy for the electrons.  This was the language in common use amongst electron lens specialists at the time, but this terminology was not in general use by those working in quantum mechanics, so it gave the impression, wrongly, that it was a particular effect that was of interest only to specialists in that field.  Nevertheless  the fringe shift was calculated correctly and paper gave a clear discussion of the consequences.  Ehrenberg and Siday \cite{wers} concluded that 
\begin{quote}
One might therefore expect wave-optical phenomena to arise which are due to the presence of a magnetic field but not due to the magnetic field itself, i.e., which arise whilst the rays are in field-free regions of space. 
 \end{quote}
	The second disadvantage was that the journal in which they published was not one of the leading journals at that time, being a publication of the British Physical Society before being taken over by the British Institute of Physics.  As a consequence, Ehrenberg and Siday have not got the recognition that they deserve from the physics community, particularly in America.  In saying this, I want to make it absolutely clear that no blame can be attached to Aharonov or Bohm.  Bohm was unaware of the original ES paper \cite{wers} when they wrote their first paper \cite{yadb59}. They came to the same conclusion independently.  In their second paper \cite{yadb61} they acknowledged the Ehrenberg and Siday paper had obtained the same results using a `semi-classical' treatment, a rather unfortunate choice of words.

The case of Ehrenberg and Siday falls neatly into what Kossel and M\"{o}llenstedt call the `Amerigo-Effekt' \cite{km93}.
Unfortunately such cases can and do happen when one is exploring a new and unexpected effect, long before the foundations of phenomenon has been properly laid.  In fact if their discussion was put into modern terms, we would see that Ehrenberg and Siday were exploring the common mathematical background shared by both optics and electron optics, namely, the symplectic group and its double cover, the metaplectic group \cite{gs84}.   The discussion of rays follows directly from the symplectic group.  In fact the rays are simply generated by a symplectomorphisms.  On the other hand, the wave properties follow from the covering group, namely, the metaplectic group.  What Ehrenberg and Siday had discovered in their own way was that the homotopy group of the covering space was non-trivial and were on the way to discovering the notion of a winding number. Alas being experimentalists, they would not have known about these advanced mathematical structures, then or even later when these techniques became more well known.

\section{Enter Aharonov and Bohm}

Ehrenberg and Siday's work remained unknown for ten years before the effect was rediscovered by Aharonov and Bohm \cite{yadb59}.  Their paper goes straight to the heart of the problem.  They note that although in classical physics the fundamental equations can always be written in terms of fields, in quantum mechanics the potentials cannot be removed from these fundamental equations, therefore this must have observational consequences.  They then ask, `What are these consequences?'

Lev Vaidman, a long time collaborator of Aharonov, told me that Yakir had spotted the vector potential producing observable effects but, did not  realise that potentials were universally considered as mere mathematical artefacts. Ah! the innocence of young research students! He went to talk with Bohm, his then supervisor, and they discussed the idea.  This discussion led them to propose an actual  experiment based, in actual fact, on Figure 1 that appeared in the Ehrenberg and Siday paper.  Their proposal was to place a shield to the right of the two slits and then to place immediately to the right of the shield, in its geometric shadow, a small long solenoid  with its axis parallel to the slits.  To ensure none of the field of the solenoid could spill out into the region of the electron paths, one could place a strip of mu-metal suitably shaped to conduct the field produced by the ends of the solenoid around the electron paths.  This ensures that the electrons move in a  field free region.  This was precisely what was done later in the beautiful experiment carried by M\"{o}llenstedt and Bayh \cite{mb61} and Bayh \cite{wb62}.  

Aharonov and Bohm were rather fortunate in that an experimentalist, Robert Chambers at Bristol University where they were working, immediately set about doing an experiment to show the effect existed \cite{rc61}.  He used a magnetic whisker and clearly demonstrated the effect.  However because of the unexpected nature of the effect, people argued that as the magnetic whisker produced an unshielded field,  the effect may be due, after all, to the field rather than the potential.  This was wishful thinking.  However the appearance of Bayh's results immediately showed that any arguments about stray fields causing the effect could be ruled out.  Since those early days a number of more refined experiments have all confirmed the effect.
The full details of all these experiments can be found in a review article by Olariu and Popescu \cite{op85}.

\section{Conclusion.}

	It is now clear that the AB effect arises directly from the Schr\"{o}dinger equation as was explained in the first Aharonov and Bohm paper \cite{yadb59}. Yet clear as the paper was, it too was received with some scepticism and even opposition to begin with.  For example, Victor Weisskopf \cite{vw61} wrote in some Brandies lecture notes, 
\begin{quote}
The first reaction to this work is that it is wrong; the second is that it is obvious.
\end{quote}
  This effect is now considered to be of great theoretic importance as it is the first example of quantum gauge phenomena.  Gauge theories have become central to the modern theory of particle interactions, spawning many examples of gauge phenomena \cite{xmq}.  The importance of the effect is reflected in a leader article published in Nature where it was proposed that Aharonov and Bohm should share the Nobel prize with Michael Berry  for their contributions to the understanding of gauge effects.  I asked Bohm for his reaction to this suggestion.  He replied that he did not think the AB effect alone was that noteworthy and added: ``After all it is only a straight forward application of standard quantum mechanics, and anyway Ehrenberg and Siday were there first!"   For me that sentence the way physics should be done with, humility, generosity and honesty.
  
  \section{Acknowledgments}
  
  I should like to thank my colleague David Butt without whose detailed reminiscences of his interaction with Ray Siday, without this background, this paper would have been a poorer report.  I am also indebted to the late John Jennings for his input and  translation of the abstract of Franz's talk given in Danzig in 1939.



\end{document}